\documentstyle[twoside,fleqn,espcrc2,epsf]{article}

\newcommand{\AmS}{{\protect\the\textfont2
  A\kern-.1667em\lower.5ex\hbox{M}\kern-.125emS}}

\newcommand{\noi}{\noindent}
\newcommand{\eq}{\begin{equation}}
\newcommand{\en}{\end{equation}}
\newcommand{\eqa}{\begin{eqnarray}}
\newcommand{\ena}{\end{eqnarray}}

\newcommand{\hg}{{\hat \Gamma}}

\newcommand{\aleq}{\mbox{}_{\textstyle \sim}^{\textstyle < }}
\newcommand{\ageq}{\mbox{}_{\textstyle \sim}^{\textstyle > }}

\newcommand{\be}{\begin{equation}}
\newcommand{\ee}{\end{equation}}
\newcommand{\bea}{\begin{eqnarray}}
\newcommand{\eea}{\end{eqnarray}}
\title{
\vspace{-1.8cm}
\hbox{}
{\small NOVEMBER 1994} \hfill 
\hbox{}                \hfill {\small HU BERLIN--IEP--94/27~}   \break
                                                              \break
Pseudoscalar correlators and the problem of the chiral
limit in the compact lattice QED with Wilson fermions
\thanks{TALK GIVEN AT THE LATTICE '94 INTERNATIONAL SYMPOSIUM
       LATTICE FIELD THEORY, BIELEFELD, GERMANY,
       SEPTEMBER 27 -- OCTOBER 1, 1994}
\thanks{Work supported by the Deutsche
Forschungsgemeinschaft under research grant Mu 932/1-3 }
}

\author{A.~Hoferichter,
V.K.~Mitrjushkin\thanks{Permanent adress:
JINR, Dubna, Russia}
and
M.~M\"uller-Preussker \\
\vspace{0.3cm}
Institut f\"{u}r Physik, Humboldt-Universit\"{a}t,
10099 Berlin, Germany
}

\begin{document}

\begin{abstract}
The phase diagram for the compact lattice QED with Wilson fermions
is shown.
We discuss different methods for the calculation of the 'pion' mass
$m_{\pi}$ near the chiral transition point $\kappa_c(\beta )$.
\end{abstract}

\maketitle

\section{Phase diagrams}

The standard Wilson lattice action
$ S_{W\! A}$ for $4d$ compact $~U(1)~$ gauge
theory (QED) is

\eq
 S_{W\! A} = \beta \cdot S_{G}(U) + S_{F}(U, {\bar \psi}, \psi) ,
                                              \label{wa}
\en

\noindent where $~S_{G}(U)~$ is the plaquette
(Wilson) action for the pure gauge  $~U(1)~$ theory,
and

\eq
S_{F} = {\bar \psi} {\cal M} \psi
\equiv {\bar \psi} [ \hat{1} - \kappa \cdot Q(U) ] \psi
\en

\noindent is the fermionic part of the action.

In our calculations we also used the modified (compact) action
$S_{M\! A}$ with Wilson fermions

\eq
 S_{M\! A} = S_{W\! A}(U) + \delta S_{G}(U)~,
                                              \label{ma}
\en

\noindent where the additional term $~\delta S_{G}~$ is introduced
to suppress some lattice artifacts, i.e., monopoles
and negative plaquettes.

The phase diagram for the standard compact theory (with $S_{W\!A }$)
is shown in Fig.1a  \cite{qed1} (the preliminary variant was in
\cite{qed2}).
The line from $(\beta_1 ;\kappa_1)$
to $(\beta_2 ; \kappa_2)$ was found to be a 1$^{\rm{st}}$ order
transition line which is driven by monopole condensation in the
confinement phase. The order of the phase boundary between
the confinement and the 4$^{\rm{th}}$ phase (i.e. the
transition line from $(\beta_2 ; \kappa_2)$ to $(\beta=0 ;1/4)$~)
still remains open. A straightforward way to tackle this problem
seems impossible, since standard fermionic bulk
observables like $\langle \bar{\psi} \psi \rangle$
are ill--defined due to permanent occurence
of near--to--zero eigenvalues
of the fermionic matrix in the 4$^{\rm{th}}$ phase. \\
\noi For $\kappa < \kappa_1$ and $\beta \simeq \beta_0$ the same
phase transition as in the pure gauge theory is observed ($\beta_0 \simeq 1$
denotes the critical copling in the pure gauge theory). Time
histories of the plaquette exhibit metastable states, which are
probably due to monopole loops wrapping around the
torus whereas no metastabilities occur for $\bar{\psi} \psi$.
Therefore we cannot establish a 1$^{\rm{st}}$ order phase transition in
this part of the phase diagram.\\
The 'horizontal' line
from $(\beta_1;\kappa_1)$ to $(\beta=\infty ;\kappa_c=1/8)$ which
separates the Coulomb phase from the 3$^{\rm{rd}}$ phase
corresponds to the chiral transition in this theory and is presumably
a higher order phase transition line.
For determining the chiral transition the variance
of the pion norm $~\sigma^2(\Pi )~$ turned
out to be a sensitive parameter~:
\eq
\sigma^2(\Pi) \sim V \cdot \frac{1}{N}
\sum_{i}^{N} \Bigl( \Pi_{i} - \overline{\Pi} \Bigr)^2~,
\quad
\overline{\Pi} \equiv \frac{1}{N} \sum_{i}^{N} \Pi_{i}.
\en
\noi where for the $~i^{\rm{th}}~$
configuration the value of $\Pi_{i}$ is given by
$1/4V \cdot \sum_{x y}
\mbox{Sp} \Bigl( {\cal M}^{-1}_{~x y} \gamma_{5}
{\cal M}^{-1}_{~y x} \gamma_{5} \Bigr)~$,
$V$ is the number of sites,
and $N$ is the number of measurements. (Sp denotes the
trace with respect to Dirac--indices). Near the chiral transition
$~\sigma^2(\Pi )~$ seems to exhibit the characteristic
features of a finite size scaling behaviour
of a singular point on finite volume systems.

\begin{figure}[htb]
\begin{center}
\vskip -1.2truecm
\noi
\leavevmode
\hbox{
\epsfysize=300pt\epsfbox{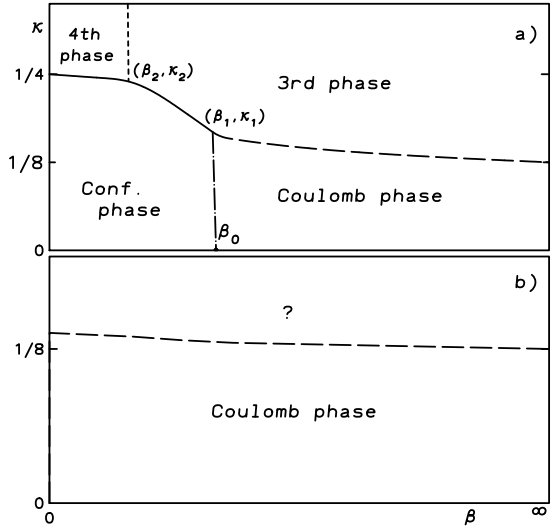}
     }
\end{center}
\vskip -3.5truecm
\caption{Phase diagram in the $(\beta,~\kappa )$--plane
for dynamical fermions for  the standard Wilson action WA ({\bf a})
and for the modified action MA ({\bf b}) valid for $~\kappa \aleq 0.3~$.}
\label{fig:pot}
\vskip -0.6truecm
\end{figure}

In case of the modified theory (eq.(\ref{ma})) the situation
drastically changes in comparison with the standard theory (see Fig.1b).
Just two phases separated by a 'horizontal' chiral transition
line $\kappa_c(\beta)$ survive. There is no sign for
a 1$^{\rm{st}}$ order phase transition even at strong coupling.
The average plaquette and the fermionic condensate, respectively,
behave smoothly with varying $~\kappa~$ at fixed $~\beta~$, no
metastabilities are observed.
For greater detail and more discussion on the phase structure of
both theories see \cite{qed1}.

\section{'Pion' mass near $~\kappa_c~$}

The standard  choice of the estimator for the effective mass of the
pseudoscalar particle (for simplicity we'll call it
the effective 'pion' mass)
$~m^{ef\! f}_{\pi}(\tau ) \equiv m_{\pi}(\tau )~$ is

\eq
m_{\pi} (\tau ) =
- \ln \frac{\hg (\tau +1)}{\hg (\tau )} ~;
\quad (\mbox{at}~~~~N_{\tau} =\infty)
                 \label{pmass1}
\en

\noindent where $\hg (\tau )$ is the pseudoscalar (zero--momentum)
correlator
\eq
\hg (\tau ) = \frac{1}{N} \cdot \sum_{i=1}^N \Gamma_i (\tau ) ~,
                 \label{corr1}
\en

\noindent At large enough
$\tau $ the dependence of $m_{\pi}(\tau )$ on $\tau $ should
disappear (at least, in
the case of the mass gap), and the resulting plateau gives the
true mass $m_{\pi}$.

We will consider the approach to the chiral
limit, i.e. $\kappa \rightarrow \kappa_c(\beta)$ in the confinement
phase ($\beta < \beta_0$) of the quenched standard theory.
The well--known problem in QED and QCD with
the calculation of $m_{\pi}$
(and other observables) is connected with extremely small
eigenvalues of the fermionic matrix ${\cal M}$ which appear
at $\kappa \sim \kappa_c (\beta)$.
In Figs.2a,b the values of $\Gamma_i (\tau )$ are shown on
individual configurations $i=1,..,500$ which form the correlator
corresponding to eq. (\ref{corr1}). The data are obtained from
a $12 \times 4^3$ lattice at $\beta=0$ and $\kappa=1/4$
in the quenched approximation.
The huge spikes of magnitude $\sim 10^4$ appearing on some
configurations make the calculation of the mass $m_{\pi}$ in
eq.(\ref{pmass1})
unreliable with the statistics at hand ($\sim 10^2 \div
10^4$ measurements).

\begin{figure}[htb]
\begin{center}
\vskip -2.0truecm
\leavevmode
\hbox{
\epsfysize=300pt\epsfbox{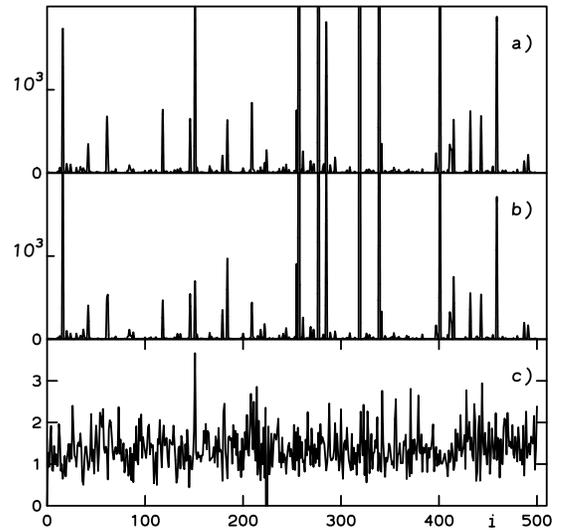}
     }
\end{center}
\vskip -3.5truecm
\caption{'Time'--histories for pseudoscalar correlators $\Gamma_i (\tau )$
at $\tau =1$ ({\bf a}),  $\tau =2$ ({\bf b}) and their ratio ({\bf c}).
$\kappa =0.25$ and $\beta =0$.}
\label{fig:history}
\vskip -0.6truecm
\end{figure}

The usual way to avoid this problem is to decrease $\kappa$ which entails
however the {\it increasing} of the mass (or $m_{\pi}/m_{\rho}$) up to
unphysical values.

A possible way to extract reliable values of $m_{\pi}$ near the
chiral transition point is based on the following observation.  The
'time'--histories for the ratios $e^{-\mu_i(\tau )} \equiv \Gamma_i(\tau
+1)/\Gamma_i(\tau )$ do {\it not} show such big fluctuations
as the $\Gamma_i(\tau)$'s themselves (see
Fig.2c).  Representing the correlator $~\Gamma_i(\tau )~$ in the form
$~
\Gamma_i(\tau ) \equiv \gamma_i \cdot f_i(\tau )~,
$
one can conclude that for every $~i^{\mbox{th}}~$
configuration the contribution of the near--to--zero
eigenvalues producing the peaks is {\it factorised out} in the
$\tau$--independent factor $\gamma_i$.
On the contrary, $~e^{-\mu_i}$ or $~\mu_i~$ are statistically
very well defined even at $~\kappa \ageq \kappa_c~$.

{}From eq.(\ref{pmass1}) one can obtain

\eq
e^{- m_{\pi}(\tau ) } = e^{-m^{\prime}_{\pi} (\tau ) } +
\frac{1}{N} \sum_{i=1}^{N} e^{-\mu_i (\tau )}  \cdot
\frac{ \delta \Gamma_i(\tau ) }{ \hg(\tau ) }
                 \label{pmass2}
\en

\noindent where

\eq
e^{-m^{\prime}_{\pi} (\tau ) } \equiv
\frac{1}{N} \sum_{i=1}^{N} e^{-\mu_i (\tau )}
\en

\noindent with
$\delta \Gamma_i(\tau ) \equiv \Gamma_i(\tau ) - \hg(\tau )$ and
$\sum_{i}^{N} \delta \Gamma_i(\tau ) \equiv 0$.
The second term in the r.h.s in eq.(\ref{pmass2}) is
supposed (for $N \rightarrow \infty$)
to give a small correction to the first term because
of the alternating factor $\delta \Gamma_i(\tau )$ in the sum.
We checked it calculating $m_{\pi}(\tau)$ and $m_{\pi}^{\prime}(\tau)$
at small enough values of $\kappa < \kappa_c$ where
the standard definition in eq.(\ref{pmass1}) works well.
In this case both definitions give the same value of the pion mass.
In Fig.3 we show the behaviour of both estimators
$m_{\pi}(\tau=5)$ and $m_{\pi}^{\prime}(\tau=5)$ as a
function of $\kappa$ at
$\beta =0$.

\begin{figure}[htb]
\begin{center}
\vskip -1.2truecm
\leavevmode
\hbox{
\epsfysize=300pt\epsfbox{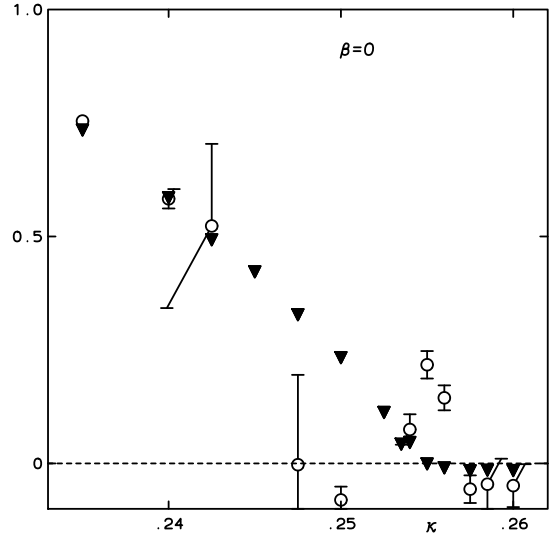}
     }
\end{center}
\vskip -3.5truecm
\caption{$m_{\pi}(\tau=5)$ (circles) and $m_{\pi}^{\prime}(\tau=5)$
(triangles) as a function of $\kappa$ at
$\beta =0$ on a $12 \times 4^3$ lattice}
\label{fig:pimass}
\vskip -0.6truecm
\end{figure}

The 'pion' mass $~m_{\pi}^{\prime}(\tau)~$ shown in Fig.3 gradually
decreases to some small value with increasing $\kappa$. Note, that
for $\kappa$'s 'far enough' from $\kappa_c$ both estimators
agree within good accuracy.
Preliminary, the transition seems to be a smooth one
(at least at $\beta=0$), while
the behaviour of the 'standard' mass $m_{\pi}(\tau)$ would
not encourage to any conclusion. A similar picture holds for the
cases $\beta=0.1$ and $\beta=0.8$ but here work is still
in progress \cite{qed4}.

\section{Conclusions}
 We have studied the phase structure of two theories with
Wilson fermions and compact action with $U(1)$ symmetry :
standard Wilson theory and a modified one with lattice artifacts
suppressed. Phase diagrams of both theories were shown.

When suppressing lattice artifacts the phase diagram changes
strongly, and there is no universality in the strong coupling
region. This resembles  the situation with staggered fermions
\cite{qed3}.

We propose another method for the calculation of the 'pion' mass
near the chiral transition point $\kappa \ageq \kappa_c(\beta )$
which hopefully solves the well--known problem arising from
near--to--zero eigenvalues of ${\cal M}$ in this
$\kappa$--region.

The behaviour of the new estimator $~m_{\pi}^{\prime}(\tau)~$ should
now be studied in the dynamical fermion case.

\end{document}